# Formal Description of Components in Operating Systems

Asen Petkov Iliev

*Abstract*— *The contemporary development of hardware components is a prerequisite for increasing the concentration of computing power. System software is developing at a much slower pace. To use available resources efficiently modeling is required. Formalization of elements, present in the material, provides the basis for modeling. Examples are presented to demonstrate the efficiency of the concept.*

## I. INTRODUCTION

Development of computer technology placed increasingly high requirements to the software, for achieving best use. The application software is itself a complex set of applications, providing support for various areas. The requirements for the software are different in terms of their sphere of application. The bottleneck of computer systems very often come out to be the realization of the components of the system software - operating system, drivers, and others and the interaction between applications and system components. Optimization of system components is a factor that is to a great extent crucial for the implementation of specialized applications. In order to achieve optimization of system components one should abide by principles that are in some respects different from those within the program. This paper presents a methodology for the formal description of the components of the operating systems. Such an approach would enable the implementation of optimization and modeling of specific situations in a computer system. 

## II. THEORETICAL BACKGROUNDS

Apart from the specific computer architecture and the specificity of the hardware implementation in the computer system, the components of an operating system can be represented as follows [1],[2]:
   - Resources - physical and logical;
   - Activities;
   - Interfaces;
   - Protocols.

The resources in an operating system are physical and logical. The presumption is that here usually virtual resources are neglected / also virtual resources can be added, having in mind that they are part of the logical resources. Natural resources are all the resources that are physically represented in the computer system - physical memory, CPU, disk, peripherals and more. Logical resources is a logical definition and is related rather to the performance of components in the system - a page in the virtual memory, a segment of a certain size, a program text, etc. From this standpoint, it can be stated that from the point of view of an operating system a program is nothing more than a certain resource, "entitled to be processed". The right of processing is explicitly granted by the designated user or the administrator of the system.

What is meant by "activities" is any identifiable impact in a system. The processing of each application is associated with processing a set of instructions in a certain way and / or in a certain order. The processing of a particular program code in the context of this definition leads to a certain activity.
There are various types of activities [2]:
   - Procedures - they can hold their own resources, as well as to benefit from global ones. They have their own identifier. The transmission of the identifier is on the basis of the *call-return* mechanism. They have a single entry point. They can also be self activated;

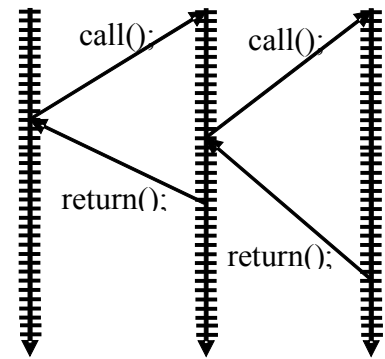

**Fig. 2.1 Diagram of the call-return mechanism**

   - Ancillary procedures - they are a generalization of the concept of procedures. They possess own local data structures, but can also benefit from the global ones. Unlike proper procedures, the ancillary ones use mechanism transfer () to enable and disable. This mechanism provides a targeted and explicit transfer of control under certain conditions. Each activated ancillary procedure is processed until it hands over the control by means of transfer (). After reactivation of an associated procedure, its processing continues from the point at which the previous transfer has implemented ();

· **Asen Petkov Iliev**, Computer Science, Burgas University "Prof. Dr. Assen Zlatarov", Burgas, Bulgaria, +359887202418., asen@asen.iliev.name.

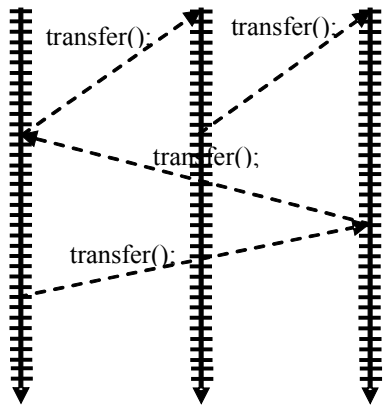

**Fig. 2.2 Mechanism transfer()**

- Processes - they are to a great extent separate and independent elements of the operating system. Each process has at least one activation identifier. A characteristic of the process is the presence of its global status and its micro status respectively. The global status of a process describes what it is currently doing - is it active, waiting, etc. The micro status of a process defines the degree to which it has been processed to a particular moment. The micro status can be determined also alternatively, on the basis of number of instructions processed until a certain moment, the number of resources at a given time, etc. Each process is handled in its own address space.

There are other forms of activities - threads, fibers, etc., but they are largely a result of a definition by manufacturing companies and can be attributed to the above. This paper will try to stick to the academic view, rather than handle with extra facts used by the companies to the implementation of market policy.

Interfaces define the static aspects of the relation between corresponding components. They describe the syntax of the relationship - what data and structures will be used and in what way. Interfaces can be management and data-oriented.

Protocols describe the dynamic aspects of the relation between corresponding components. They describe the manner of communication. Protocols are essential to describe the semantics of communication between components.

### III. FORMAL DESCRIPTION OF THE COMPONENTS

*A. Formal definition of resources*

For the sake of clarity, in the beginning it is assumed that each resource is a set elements [2], structured in the same way (memory - a collection of cells; processor - a collection of records; CD - a collection sectors, etc.). A resource in this assumption consists of multiple *M*, abstract, structured in the same way elements m, having *adr* address and corresponding address value w, ie:

$$m = (adr, w) \quad adr\ \varepsilon\ ADR,\ w\ \varepsilon\ W$$

The number of all $m\ \varepsilon\ M$ defines the current state of the resource at a certain point of time. Both the individual element m, and the constructed by the process M, are on this resources time-dependent. The symbol $Z_P(t_i)$ will be used to represent the status of the resource *M*, in a discreet point of time $t_i$:

$$Z(t) = \begin{bmatrix} w_1(t) \\ w_2(t) \\ ... \\ w_n(t) \end{bmatrix} \quad provided\ that\ ADR = \{1\ ....\ n\}$$

Moreover, each element can be associated with functions $f \in FUNC$, transforming the state of the resources:

$$f : M \to M$$
$$Z(t_{i+1}) := Z(t_i)$$

For simplification it can be assumed that *FUNC* covers all functions that could be applied to the resource.

The formal definition of a resource can be summarized as follows: a resource is an arranged four, consisting of an identifier *id*, address space *ADR*, multitude of the lending value *W* for each element of the resource and a host of features *FUNC*, which can be applied to each of the elements:

$$R = (id, ADR, W, FUNC)$$

*B. Formal definition of a process*

On the basis of the verbal description of the concept process *P*, the following formal description can be given:

$$P = (id,\ R_a,\ R_r,\ F,\ W_a(R_a))$$

Individual elements of the process P mean:

- Id -   unique identifier;

- $R_a$ -   A multitude of set aside resources (vector of the resources owned by the process);

- $R_r$ -   A multitude of resources, claimed by the process (vector of the resources, requested by the process);

- F-   sequence of instructions f (f ε F, wherein f: R -> R ');

- $W_a(R_a)$   The values of all the resources allocated during the emerging of the process.

Processes must be clearly distinguished from one another and therefore *id* is there. From its very emerging, a multitude of resources are set aside for the process. The operating system must have information about them – what type, which specifically and in what quantity. Each machine instruction in the composition of a program (other than a.e. NOP) is changing the values, of certain resources. Thus, the treatment process can be extended to a sequence and/or a set of transformations of the individual resources. The program itself is not a set of transformations. It just describes them. And last, but not least, are the allocated resources requested by the process. For example: in order to ensure a high level of security a reset of the values of the requested memory is required in advance.

### IV. APPLICATION OF FORMAL DESCRIPTION

*A. Model of the states of the process*

Due to the fact that a computer system processes a number of processes, substantially larger than the number of processors, it is necessary to establish the global states of the processes. At a certain point it is only possible to process a number of processes exactly as the number of processors. Statecharts in different operating systems differ, but these differences are not drastic from the general model that guarantees at least three states [3],[4]. On the basis of the formalization introduced, this model can be described as follows:

Active: The process possesses all of its requested resources, including a processor. Formally, this condition is described as follows:

$$|R_a| = |R_r|$$

readiness: The process possesses all of its requested resources, but with no processor. The formal description is as follows:

$$|R_a| = |R_r| - 1$$

Blocking: The process is waiting for (a) requested allocated resource (s). This state looks formally like this:

$$|R_a| < |R_r| - 1$$

Formal definitions to a great extent simplify the description of the state diagram.

### B. Formalization of a layer model of operating system

At this point, using a formal definition, an intuitive model of an operating system, consisting of m number of layers, will be shown. The S multitude of all Si layers is defined like this:

$$S = |S^i | 0 \leq i \leq m-1|$$

The model is limited by the lowest $S^0$, layer being the hardware platform and the last $S^{m-1}$, the highest in rank, which is determined by the one, creating the model (at OSI, m = 7) [5]-[7].

Two adjacent layers $S^j$ и $S^{j+1}$ are associated with two different relations:

1) The higher layer $S^{j+1}$ uses the services of the lower $S^j$:

$$S^{j+1} \nabla_d S^i$$

2) Lower layer $S^j$ controls the higher $S^{j+1}$:

$S^{j+1} \nabla_s S^i$ is used in the sense of "call", "uses"

The sequence of relations in servicing $S^{m-1} \nabla_d S^{m-2}, S^{m-2} \nabla_d S^{m-3}, ... S^1 \nabla_d S^0$ constitute the hierarchy of service. Each $S^j$ layer has resources. The multitude of resources of each layer is represented as follows:

$$R^i = |R^i_j | 0 \leq j \leq n-1|$$

Each layer may comprise of a different number of resources (hence the parameter n). Each resource is consolidated only with a particular layer:

$$R^i \cap R^j = 0 \quad (0 \leq i, j \leq m-1 \wedge i \neq j)$$

From the set of all resources, available in a given layer $S^j$ the subsets can be defined, which in a certain aspect can be tested. Out of them a control aggregation can be formed:

$$C^i_j \subseteq R^i$$

This control unit presents a particular activity, but it belongs to the same layer from which the resources come. Due to the fact that each activity has a time-limited action (limited life) it can be defined as a temporary resources aggregation (temporary aggregation). Example: Based on a program code, memory and identifier a process is created. The Possible (theoretically, of course) number of activities will be characterized by all subsets of $R^i$, ie the power of $R^i$:

$$|C^i_j| = \prod (R^i)$$

It is obvious that not every free coverage of resources in a certain layer can form a meaningful activity. The actual formation of the activity based on a set of resources of a particular layer $S^i$ is taken from the function g, which is closely localized in the lower layer:

$$g^{i-1} : |R^i| \Rightarrow C^i \quad (i \geq 1)$$

After applying the so-defined function, the passive aggregation of resource becomes active. For example: The system call fork () (actually, fork () is corefunction, an element of the Main Library, which deals with the functionality of the system call sys_fork (), but the detailed research of system calls is beyond the scope of this paper [8]) .

The aggregation of resources of a particular layer i is not always a prerequisite for the occurrence of activity. More precisely, based on the aggregation of resources, the result may be the emerging of a new resource rather than an activity. For example: Based on a number of blocks of physical memory, you can create a logical memory. This " emerging " of a resource at a higher level could be defined by another function f, as follows:

$$f^i : |R^i| \Rightarrow R^{i+1}$$

The function $f^i$ creates a resource at the layer i +1 based on the resource from the layer i. Such a function is to be localized both in the two layers.

### V. GUIDELINES FOR FUTURE RESEARCH

The formalization, proposed in this paper could be applied in modeling processes in real-time systems, with applications in robotics. Because hardware is developing at a faster rate than programming, even though today's computers are not fundamentally different from those of 1960 year, there is a visible trend in the application of universal operating systems in specific control platforms - for example Linux or Windows. It is not kept secret that the trend of "free software", killing valuable researches, requires the use of universal operating systems, although they are "unprepared" for the management of the real time type. This approach requires modeling of complete applications in robotics. An interesting application could be the modeling of an aggregation of interfaces for complex use by blind users [9],[10].

**Asen Petkov Iliev**, Department Computer Science, Burgas University "Prof. Dr. Assen Zlatarov", Burgas, Bulgaria, +359887202418., e-mail: asen@asen.iliev.name .